\documentclass[preprint,11pt,aps,prl,showpacs]{revtex4}
\usepackage{graphicx}
\usepackage{amstext,amsfonts,amsbsy,amssymb,bbm}
\usepackage{array,multirow}
\usepackage[pageshow]{supertabular}
\usepackage{color}
\usepackage{setspace}

\def\Tr{\mathop{\rm Tr}\nolimits}

\newcommand {\be}[1]{\begin{eqnarray} \mbox{$\label{#1}$}  }
\newcommand{\ee}{\end{eqnarray}}

\newcommand{\pref}[1]{(\ref{#1})}

\newcommand{\iden}{\mathbbm{1}}

\newcommand{\nn}{\nonumber\\}

\newcommand{\bs}{\boldsymbol}

\newcommand{\pd}{\partial}

\renewcommand{\v}[1]{{\bf #1}}

\newcommand{\cD}{ {\cal D} }

\newcommand{\cP}{ {\cal P} }

\newcommand{\cS}{ {\cal S} }

\newcommand{\gD}{ {\Delta} }
\newcommand{\gf}{\phi}

\newcommand{\gs}{ {\sigma} }

\renewcommand{\ge}{ {\epsilon} }
\newcommand{\gl}{ {\lambda} }
\newcommand{\gr}{\rho}

\newcommand{\complex}{\mathbb{C}}

\begin{document}

\title{Unextendible product bases and extremal density matrices with positive partial transpose}
\author{Per {\O}yvind Sollid$^a$, Jon Magne Leinaas$^a$ and Jan Myrheim$^b$}
\affiliation{$(a)$ Department of Physics, University of Oslo, N-0316 Oslo, Norway\\
$(b)$ Department of Physics,
Norwegian University of Science and Technology,
N-7491 Trondheim, Norway}

\date{April 7 2011}
\begin{abstract}
In bipartite quantum systems of dimension $3\times 3$ entangled states that are positive under partial transposition (PPT) can be constructed with the use of unextendible product bases (UPB). 
As discussed in a previous publication all the lowest rank entangled PPT states of this system seem to be equivalent, under $SL\otimes SL$ transformations, to states that are constructed in this way. Here we consider a possible generalization of the UPB constuction to  low-rank entangled PPT states in higher dimensions. The idea is to give up the condition of orthogonality of the product vectors, while keeping the relation between the density matrix and the projection on the subspace defined by the UPB. We examine first this generalization for the $3\times 3$ system where numerical studies indicate that one-parameter families of such generalized states can be found. Similar numerical searches in higher dimensional systems show the presence of extremal PPT states of similar form. Based on these results we suggest that the UPB construction of the lowest rank entangled states in the $3\times3$ system can be generalized to higher dimensions, with the use of non-orthogonal UPBs. 

\end{abstract}

\maketitle

\section{Introduction}

One of the most striking features of quantum mechanics is found in the concept of entanglement. Considerable efforts have been made in understanding its properties and usefulness, with several different approaches taken \cite{Horodecki09}. One approach has been to study the geometrical structure of different convex sets of hermitian matrices that are related to subsets of quantum states with entanglement \cite{Kus01,Verstraete02,Pittenger03,LeinaasMyrheim06,Bengtsson06,Lewenstein01}. 
There are two convex subsets of the full set of density matrices, denoted  $\cD$, that are particularly important in this discussion. One is the set $\cS$ of non-entangled, or separable, states, and the other the set $\cP$ of density matrices  that remain positive under partial transposition with respect to one of the  subsystems. The states of $\cP$ are  commonly referred to as PPT states.  The non-entangled states are always PPT, so we have the set-theoretical relations $\cS\subset\cP\subset\cD$.

In two previous papers we have examined the properties of entangled density matrices that are PPT, which means that they are contained in $\cP$ but not in $\cS$ \cite{LeinaasMyrheim10a,LeinaasMyrheim10b}. These states which are known to have {\em bound} entanglement are intrinsically interesting \cite{Horodecki98}, but they are also interesting through the information they give about the set of  entangled states. In fact, to establish whether a given density matrix is entangled or separable is usually a difficult problem, but only if the density matrix is PPT. This is so, since any state which is not PPT - a property that can easily be verified -  is entangled.

The states studied in \cite{LeinaasMyrheim10a,LeinaasMyrheim10b} were found by performing systematic numerical searches for PPT-states with specified ranks for the density matrix $\gr$ and for its partial transpose $\gr^P$. Such searches were performed for several bipartite systems of low dimensions, and in all cases, when the ranks were sufficiently low, the states were not only entangled but were typically extremal PPT states. An interesting result was that the lowest rank states of this type, with full local ranks, were found to have very similar properties, independent of their dimension. In particular they were all found to be extremal PPT states with a finite and complete set of product vectors in their kernel, and no product vectors in their image.

For a bipartite system of dimension $3\times 3$ there is a special subset of these lowest rank entangled PPT states that can be constructed by the use of {\em unextendible product bases} (UPB) \cite{Bennett99}. The method provides a way to  combine vectors in the two subsystems to form an orthogonal product basis that spans a five-dimensional subspace in the full Hilbert space, and which cannot be extended to include other product vectors orthogonal to the five states. The corresponding density matrix is constructed, up to a normalization, as the projection on the space orthogonal to the product vectors of the UPB. With the use of the numerical method described in \cite{LeinaasMyrheim10a} a large number of more general entangled PPT states of rank 4 were generated, and it was shown that all of them, in a specific sense, were equivalent to density matrices that could be produced by the UPB construction. The equivalence classes were further shown to be parameterized by four real parameters, and the numerical results were interpreted as evidence for the conclusion that all entangled PPT states of rank 4 in the $3\times 3$ system are covered by this parametrization.

These results have motivated the present work, where we investigate a possible generalizion of  the UPB construction which makes it applicable to systems of dimensions higher than $3\times 3$. A simple copy of the UPB construction seems not possible, since in higher dimensions the orthogonality requirement is too demanding. Instead we focus on other properties of the construction. In the $3\times 3$ system the projection on the five-dimensional subspace spanned by the orthogonal product vectors of the UPB can be written as a sum of one-dimensional projections defined by the product vectors, and the corresponding density matrix is defined as the projection on the orthogonal subspace. The partial transpose of the density matrix takes the same form, when expressed in terms of a related, conjugate UPB. In the present paper we study density matrices of the similar projection form, but without the requirement of orthogonality. Such states can be found in the $3\times 3$ system, and we examine these states in some detail. We further focus on the question whether such generalized states can be found in systems of dimension higher than $3\times 3$.

The result is that we find such states in all the  bipartite  systems we have been able to examine. In addition to the $3\times 3$ system this applies to systems of dimensions 
$3\times n$ with $n$ taking values up to $6$, and we have further examined the $4\times 4$, $4\times 5$ and $5\times 5$ systems. The matrices we find are all of the projection form referred to above, with generalized, non-orthogonal UPBs in their kernel, and the partial transposed matrices are all found to have the same form. 

The organization of the paper is as follows. We first discuss in some detail the generalized  UPB construction for the lowest rank PPT states of the $3\times 3$ system. 
These states have rank 4 both for the density matrix $\gr$ and its partial transpose $\gr^P$, and we thus refer to them as $(4,4)$ states. We illustrate the generalization by examining a special, symmetric case where the vectors of the UPB form a regular icosahedron, and find a one-parameter set of equivalent states that correspond to a linear deformation of the icosahedron. One of these states has an orthogonal UPB in its kernel, while for the general case the UPB is non-orthogonal.

We next describe a numerical method to search for more general matrices, with less symmetric UPBs. The method specifies these density matrices $\gr$ to be projections with rank equal to 4 and to have a positive partial transpose. All matrices that we find in this way are quite remarkably not only projections, but have the form we refer to in the generalized UPB construction. Furthermore, the partial transpose $\gr^P$ has the same form when expressed in terms of the associated, conjugate UPB. All density matrices found in this way are extremal PPT states.

For the $3\times 3$ system we have previously found numerically that generic extremal PPT states of rank $(4,4)$, found by the method described in \cite{LeinaasMyrheim10a}, can be transformed by product transformations to a projection form with orthogonal UPBs in their kernels \cite{LeinaasMyrheim10b}. Here we further investigate whether such states can be transformed to a projection form with more general UPBs in their kernels, dropping the orthogonality requirement on the transformed UPBs. The results show 
that this is the case, and furthermore strongly indicate that the projections belong to one-parameter classes of equivalent projections, where density matrices defined by the {\em orthogonal} UPB construction are special cases.

For the higher-dimensional systems the method we use to generate density matrices of projection form with specified ranks also works well. When the rank is chosen to coincide with the rank of the lowest rank extremal PPT states, as discussed in \cite{LeinaasMyrheim10a}, we find matrices with the same properties as in the $3\times 3$ system. The projection can thus be expressed in terms of a UPB in the kernel of the matrix, and the partial transpose has the same rank and structure when expressed in terms of its associated UPB. However, our further numerical studies indicate that a generic entangled PPT state of this rank cannot be transformed, by a product transformation, into a the projection form. In this respect the higher-dimensional systems seem to be different from the $3\times 3$ system.

We end with a summary and with a discussion of some of the questions that are left for further research.

\section{Lowest rank extremal PPT states in the $3\times3$ system}

The construction of entangled PPT states of rank $4$ in the $3\times 3$ system, with the help of orthogonal UPBs, was originally discussed in \cite{Bennett99}. A condition was there given for choosing a set of five vectors $\{\phi_k\}$ in subsystem A and a corresponding set of vectors $\{\chi_k\}$ in subsystem B so that the set of product vectors $\{\psi_k=\phi_k\otimes \chi_k\}$ would define an orthogonal UPB. From this set of product vectors a density  operator of rank 4 could be constructed as a (normalized) projection operator in the following way
\be{projec}
\gr={1\over 4}(\iden -\sum_{k=1}^5 \psi_k\psi_k^\dag) 
\ee
This form for $\gr$ implies that the partial transpose $\gr^P$ will have the same form when expressed in terms of a {\em conjugate} UPB, defined as $\{\tilde\psi_k=\phi_k\otimes \chi_k^*\}$, which is also a set of orthogonal product vectors. (The complex conjugation of the second factor implies that the partial transposition is performed with respect to subsystem $B$.)

The important point is that the density matrix $\gr$, defined by the above expression, is necessarily entangled and PPT. It is entangled since there is no product state in the image of the matrix, and it is positive since it is proportional to a projection, which has only non-negative eigenvalues. For the same reason also the partial transpose is positive, and thus follows the PPT property of $\gr$. The density matrices defined by this construction form a proper subset of the entangled PPT states of rank 4 since the generic state of this type, according to our previous studies, will have a non-orthogonal UPB in its kernel \cite{LeinaasMyrheim10a}.

The idea is now to relax the condition of orthogonality, but to keep other properties of the UPB construction. We examine this generalization first in the $3\times 3$ system, but will subsequently examine the corresponding generalization for higher dimensional systems. The main condition is that the rank $4$ density operator should be proportional to a projection operator, which we write as 
\be{genprojec}
\gr={1\over 4}(\iden -Q)\,,\quad Q^2=Q
\ee
where $Q$ is assumed to be of the form
\be{qpro}
Q=5\sum_{k=1}^6 p_k \psi_k\psi_k^\dag
\ee
with $\{\psi_k=\phi_k\otimes \chi_k\}$ as a set of product vectors that defines a generalized UPB. The coefficients $p_k$ define an unspecified set of real parameters, not necessarily all positive,  with $\sum_{k=1}^6p_k=1$.

Note that the sum over the product vectors now runs from 1 to 6. The reason for this is the following. The five dimensional subspace spanned by the UPB will always include a total of 6 product vectors \cite{LeinaasMyrheim10a}. When five of the product vectors are orthogonal, the 6th product state which is a linear combination of the other 5, is simply not included in the definition of $\gr$. We may view this as a special case of \pref{genprojec} and \pref{qpro}, with $p_6=0$. However, when there is no subset of the product vectors that is orthogonal it seems natural to define the generalization so that all product vectors in the five dimensional subspace are included, as we have done above.

The density matrix $\gr$ defined by \pref{genprojec} and \pref{qpro} is clearly entangled, since it has no product vector in its image. Furthermore, the partial transpose $\gr^P$ has the same form as $\gr$, when the product vectors $\psi_k$ are replaced by the conjugate vectors $\tilde\psi_k=\phi_k\otimes \chi_k^*$, so that
\be{genprojec2}
\gr^P={1\over 4}(\iden -Q^P)
\ee
with $Q^P$ given by
\be{qpro2}
Q^P=5\sum_{k=1}^6 p_k \tilde\psi_k\tilde\psi_k^\dag\,
\ee
Also in this case the vectors $\tilde\psi_k$ form a generalized UPB. However, it is not obvious from the above expressions that the rank of $Q^P$ is the same as the rank of $Q$, and neither that $\gr^P$ is positive. 
We only here note that in earlier numerical studies of the lowest rank extremal PPT states, we find that $\gr$ and $\gr^P$ appear generically with equal ranks (in the $3\times 3$ system as well as in higher dimensions)\cite{LeinaasMyrheim10a}. This implies that, for these states, if $\gr$ is on projection form so is $\gr^P$, as we shall show explicitly in a later section.  In the following we shall simply assume that the generalized UPB construction will treat $\gr$ and $\gr^P$ in a symmetric way with respect to the conditions \pref{genprojec} and \pref{qpro}, and thereby secure that $\gr$ is both entangled and PPT.

Since we lack  an explicit prescription for choosing vectors $\gf_k$ and $\chi_k$ of the subsystems so that the product vectors $\psi_k=\gf_k\otimes\chi_k$ define density matrices $\gr$ and $\gr^P$ that satisfy the above conditions, we have focussed instead on the question if we can, by numerical searches, find states that satisfy these criteria. This we have done first for the $3\times 3$ system and subsequently for higher dimensional systems. In the $3\times3$ system we can, however, make an explicit construction of such states as a special case, and we shall discuss that case in the next section. 

\section{A special case: The icosahedron}
An especially symmetric case, with non-orthogonal product states, is formed by the icosahedron construction described here. The vectors $\phi_k$ and $\chi_k$ of the subsystems that combine into the product vectors of the generalized UPB are in this case all chosen to be real. The six vectors $\gf_k$ define the six symmetry axes of a regular icosahedron that pass through its twelve corners. A particular choice of (non-normalized) vectors is specified by the following sets of Cartesian coordinates
{\samepage
\be{veccoord}
&&\phi_1=(-\gf,0,1)\,,\;\; \;\phi_2=(-1,\gf,0)\,,\;\; \;\;
\phi_3=(1,\gf,0)\,,\;\; \nn
&&\phi_4=(\gf,0,1)\,,\;\; \;\;\;\;
\phi_5=(0,-1,\gf)\,,\;\;\;\; \phi_6=(0,1,\gf)
\ee
}
with $\gf=(\sqrt 5 +1)/2$ as the golden ratio. These are also the coordinates of six of the corners of the icosahedron, when the center is located at the origin. 

The second set of vectors, $\{\chi_k\}$, is chosen as the same as set \pref{veccoord}, but in order to form the correct combinations $\psi_k=\gf_k\otimes \chi_k$, the $\gf_k$ and $\chi_k$ vectors have to be differently ordered. A particular choice is $\chi_k=\gf_{(2k+4)\mbox{\tiny mod}5}$ for $k=1,...,5$ and $\chi_6=-\gf_{6}$, but in total there are 60 acceptable orderings, related to the 60 rotational symmetries of the icosahedron. With all these orderings the six product vectors span a five-dimensional  subspace of the Hilbert space.

The vectors \pref{veccoord} define six equiangular lines, which means that the scalar product between all pairs of vectors are equal up to a sign. With normalized vectors the scalar products are
\be{scalarprod}
g_{kl}^A\equiv \gf_k^\dag \gf_l=\pm {1\over\sqrt 5}\,,\quad 
g_{kl}^B\equiv \chi_k^\dag \chi_l=\pm {1\over\sqrt 5}\,,\quad k\neq l
\ee
For the product vectors $\psi_k=\gf_k\otimes \chi_k$ the scalar products are also equal up to a sign, and by choosing the particular ordering of the $\chi_k$, referred to above,  we obtain
\be{fullprod}
g_{kl}=g_{kl}^A \,g_{kl}^B= -{1\over 5}  ~\forall~ k\neq l,~~k,l=1,...,6
\ee

We now consider the following operator
\be{qop}
Q={5\over 6} \sum_{k=1}^6 \psi_k\psi_k^\dag
\ee
which is of the form \pref{qpro} with all six coefficients equal, $p_k=1/6$. The condition that it should define a projection, $Q^2=Q$, can be written as
\be{linsup}
\psi_k=5\sum_{l\neq k} g_{lk} \psi_l\,, \quad g_{lk}=\psi_l^\dag \psi_k
\ee
and this should be satisfied for all $k$. When $g_{kl}$ is chosen as in (\ref{fullprod}) it gives the symmetric condition
\be{symcond}
\sum_{k=1}^6 \psi_k=0
\ee
and it is straight forward to check that this is equation is satisfied for the product vectors of the regular icosahedron.

The product vectors $\{\psi_k\}$ define a generalized UPB. Thus, there is no product vector orthogonal to the set, as can easily be checked, and the vectors are non-orthogonal. The set defines an entangled PPT state in the form of the density operator
\be{dens}
\gr={1\over 4}(\iden -{5\over 6} \sum_{k=1}^6 \psi_k\psi_k^\dag)
\ee
It is entangled since there is no product vector in its image, and it is PPT since the vectors are all real, and the partial transposition thus leaves the density operator invariant, $\gr^P=\gr$.

\begin{figure}[h]
\begin{center}
\includegraphics[width=9cm]{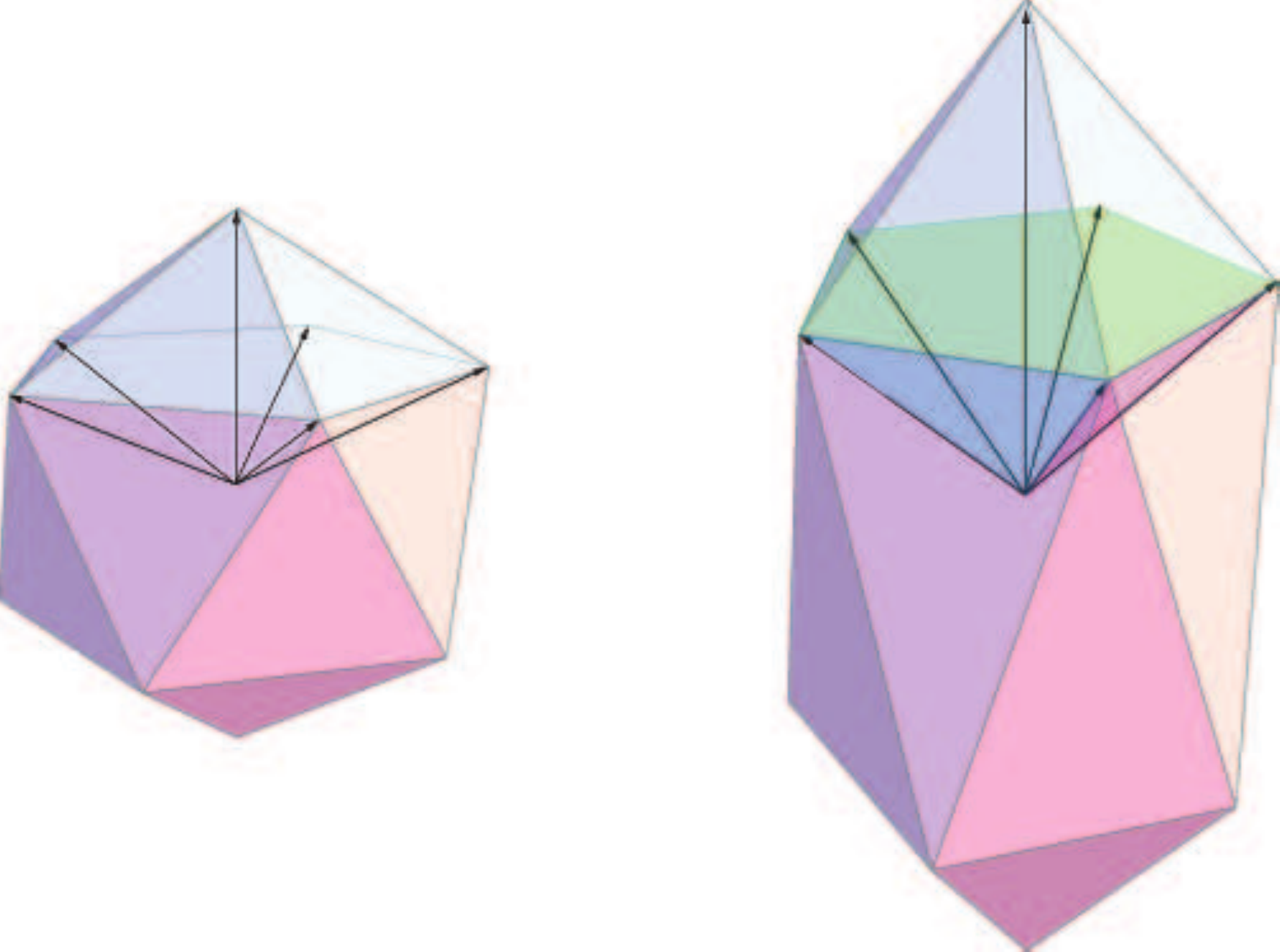}
\end{center}
\caption{\small The icosahedron construction. A regular icosahedron, shown to the left, defines a set of six equiangular lines through the twelve corners. A corresponding set of six vectors along these lines are shown in the figure. Tensor products of six pairs of such vectors define the vectors of the UPB, as explained in the text, and all of these appear with equal weight in the construction of the corresponding entangled PPT state. To the right a stretched icosahedron is shown where five of the vectors are used to define an orthogonal UPB. These vectors point along edges of a regular pyramid with a pentagon as base, which appears upside down in the figure. This construction has previously been referred to as the Pyramid \cite{Bennett99}. The corresponding UPB defines an entangled PPT state where the five product states appear with equal weight, and where  the sixth vector does not appear.    \label{Icosahedron}}
\end{figure}

States that are related by non-singular product transformations
\be{equiv}
\gr'=S\gr S^\dag\,,\quad S=S_A\otimes S_B
\ee
where $A$ and $B$ refer to the two subsystems, 
all have the same characteristics as being separable or entangled. They also share the property of being positive or not under partial transposition and they have the same rank.
This is trivially the case for unitary product transformations, but it is also true for non-unitary transformations, in which case the transformed matrix $\gr'$ should further be normalized to unity. The operators $S_A$ and $S_B$ can be restricted, without loss of generality, to be unimodular, and for this reason we refer to the relation as $SL\otimes SL$-equivalence (or simply $SL$-equivalence) \cite{LeinaasMyrheim10b}. 
For the density operator \pref{dens} there is a subset of such $SL$-equivalent states that can be written in the form of projections \pref{qpro}. Clearly unitary product transformations will leave this form invariant, but it is of interest to note that there is a one-parameter set of non-unitary transformations that also leaves the projection form invariant. As opposed to the unitary transformations these transformations will change the coefficients $p_k$ in the expansion \pref{qpro}.

The non-unitary $SL$-transformations are of the symmetric form $S=S_A\otimes S_B$ with $S_A=S_B$, where $S_A$ rescales the length in the direction of one of the symmetry axes of the icosahedron. In the following we choose this to be the direction of $\gf_6$, so that the transformation squeezes or elongates the icosahedron along this axis. This can be described in terms of a streching parameter $\gl$, so that the unit vectors pointing towards the corners of the deformed icosahedron are
\be{deform}
\gf_k(\gl)=N(\gl) \left(\gf_k+(\gl -1)(\gf_6^\dag\gf_k)\,\gf_6\right) \,,\quad k\ne 6\,;\quad \gf_6(\gl)=\gf_6
\ee
with $N(\gl)$ as a normalization constant. For symmetry reasons, the coefficients $p_k(\gl)$ are all equal for $k\neq6$. It is straight forward to show that
\be{p6}
p_6(\gl) = \frac{4 + 2\gl^2 - \gl^4}{20 + 10\gl^2}; ~~~p_k(\gl)=\frac{(1-p_6(\gl))}{5} ~~\forall~~ k\neq6
\ee

The stretching parameter is chosen with $\gl=1$ for the regular icosahedron with $p_k=1/6$ for $k=1,...,6$, in which case all product vectors contribute equally. When $\gl=0$ the squeezed icosahedron collapses to a plane, at which point the set of product vectors cease to be a UPB since the vectors $\gf_k$ for $k=1,...,5$ all lie in the two-dimensional plane orthogonal to $\gf_6$. For the particular value $\gl=\sqrt{2\gf}$ the five product vectors, with $k=1,2,...,5$, become orthogonal, and thereby define an orthogonal UPB. This is the Pyramid construction discussed as a particular case of a UPB construction in \cite{Bennett99}. In this case the coefficients are $p_k=1/5$ for $k\neq 6$ and $p_6=0$. The relation between this set and the stretched icosahedron is illustrated in Fig.\ref{Icosahedron}.

\section{The general case of rank $(4,4)$ entangled PPT states}
We focus now on general entangled PPT states of rank $(4,4)$ in the $3\times 3$ system, which are necessarily also extremal PPT states. In a previous study \cite{LeinaasMyrheim10a} we have numerically produced a large number of such states, and we have found that they all can be transformed, by product transformations, to states of the special form given by \pref{projec}, with an {\em orthogonal} UPB  in the kernel of the density matrix \cite{LeinaasMyrheim10b}. The question that we now discuss is whether these are only special cases, and that more general states on the projection form given by \pref {genprojec} and \pref{qpro} can be found. In order to examine this we have applied two different numerical methods to search for density matrices of the desired form.

The first method  applies search criteria which do not directly refer to the conditions \pref {genprojec} and \pref{qpro}. Instead the search is for density matrices $\gr$  with the correct rank $4$, which are PPT and of projection form.  This means that all the non-vanishing eigenvalues should be equal. 
The method is essentially the same as used in \cite{LeinaasMyrheim10a} to search for PPT states of rank $(4,4)$. Here however, we impose no restriction on the rank of $\gr^P$, only that $\det\rho^P\geq0$.  It is a linearized, iterative method which specifies a certain number of the eigenvalues of $\gr$ to vanish, and in this case also the remaining eigenvalues to be equal. We refer to \cite{LeinaasMyrheim10a} for details concerning the iterative method. A similar approach is also used in the second method, to be discussed below.

The result is that the numerical method works  well, and a large number of states which satisfy the search criteria have been found. Even if we do not introduce any explicit constraint on $\gr^P$, only the conditions that $\gr$ is PPT and is proportional to a projection, we find always $\gr^P$ also to be proportional to a projection, so that both $\gr$ and $\gr^P$ satisfy the condition \pref {genprojec}. Futhermore we find in all cases that $\gr$ (and $\gr^P$) satisfy condition \pref{qpro}, when expressed in terms of the vectors  of the UPB, even if that is not included as a condition in the searches.

To gain some more information about the type of density matrices we want to obtain, we have applied a second numerical method. It introduces a search for a product transformation which maps, if possible,  a chosen rank $(4,4)$ extremal PPT state into the form of a projection. We do not, in this search either, impose any condition on how this projection is expressed in terms of product vectors. Also this method applies a linearized, iterative approach to determine the product transformation, which we now outline.

Let us denote the initial density matrix by $\gr_0$ and the final density matrix by $\gr$. Their relation should then be of the form
\be{transform}
\gr=S \gr_0 S^\dag\,,\quad S=S_A\otimes S_B
\ee
and the condition that the transformed density matrix is a projection gives
\be{prjec}
S\gr_0 S^\dag S\gr_0 S^\dag=S\gr_0 S^\dag
\ee
(Note that the density matrix $\gr$ is then not normalized to unity.) The transformation $S$ is assumed to be non-singular, and the condition can then be simplified to
\be{transform2}
\gr_0 S^\dag S\gr_0 =\gr_0 
\ee

The unknown transformation matrix $S$ we parametrize by expressing $S_A$ and $S_B$ as linear combinations of a complete set of $3\times 3$ hermitian matrices.  The corresponding product transformations are parametrized by 17 real parameters $\mu_k$, which we interpret as components of a vector $\bs\mu$. Furthermore we express the rank 4 matrices as 16-component vectors in the space of hermitian operators within  the image of $\gr_0$. Written as a vector equation, Eq. \pref{transform2} has the form
\be{veceq}
\bs F(\bs\mu)= \bs\gr_0
\ee

Assume now that $\bs\mu'$ is a trial vector that gives an approximate solution to Eq. \pref{veceq}. We write the deviation from the true solution as $\gD \bs\mu=\bs\mu-\bs\mu'$, and treat this as a perturbation. To first order in $\gD\bs\mu$ the equation reads in matrix form
\be{lin}
\bs B \gD\bs\mu= \bs\gr_0-\bs F
\ee
with $\bs B$ as a real, non-quadratic $16\times 17$  matrix with elements $B_{kn}=\pd F_k/\pd \mu_n$, and where both $\bs F$ and $\bs B$ are evaluated with the trial vector $\bs\mu'$. By multiplying with the transposed matrix $\bs B^T$ and introducing the positive, real symmetric matrix $\bs A =\bs B^T\bs B$, as well as $\bs a=\bs B^T (\bs\gr_0-\bs F)$, the equation can be written in the form
\be{veceq2}
\bs A  \gD\bs\mu= \bs a
\ee
where $\v a$ and $\bs A$ are determined by the trial vector $\bs\mu'$ and $\gD\bs\mu$ is the unknown to be determined by the equation. Written in the form \pref{veceq2} the equation is well suited to be solved numerically by the conjugate gradient method \cite{Golub83}.

An iterative approach is used to find a solution of the original problem. A starting point $\bs\mu_1$ is chosen for the trial vector, and this determines the initial versions of $\v a$ and $\bs A$. Eq.~\pref{veceq2} is then solved numerically to give a first solution $\gD\bs\mu_1$. The  trial vector is then updated with $\bs\mu_2=\bs\mu_1+\gD\bs\mu_1$, this vector is used to improve $\v a$ and $\bs A$, and a new improved solution $\gD\bs\mu_2$  of  \pref{veceq2} is found. If repeated iterations of this procedure leads to convergence, in the sense $\gD\bs\mu\to 0$, the limit value of $\bs\mu$ gives a solution to the original problem \pref{veceq} and \pref{transform2}. 

A possible problem with this method is that the solution we find may correspond to a singular transformation matrix $S$, which will not give a density matrix $\gr$ with correct properties. However, for the $3\times 3$ system, by repeatedly applying the method with different starting points, we have found that  usually the convergence of the method is quite rapid, and the solution that we find corresponds to a non-singular product transformation. 

The result from applying the above method to the rank $(4,4)$ states of the $3\times 3$ system is that for a large number of initial density matrices that we have used, we can in all cases transform the density matrix by a non-singular product transformation to a matrix with the form of a projection. In all cases the density matrices $\gr$ that we find have the form specified by  \pref {genprojec} and \pref{qpro}, when expressed in terms of the (non-orthogonal) UPB in the kernel of the density matrix. Furthermore, also the partially transposed density matrices $\gr^P$ are,  in all cases, proportional to projections, and can be written in the form given by  \pref {genprojec} and \pref{qpro}. This happens even if only $\gr$ is required to have the form of a projection in the numerical search.

By keeping $\gr_0$ in \pref{transform2} fixed and choosing repeatedly different starting points $\bs\mu_1$ in \pref{veceq} for the first iteration, we get different results for the the density matrix $\gr$ determined by the above method. This indicates that there is a large number of different states on projection form within a given set of $SL$-equivalent $(4,4)$ states. In fact there are several indications that there is a one parameter set of such states, which can be related by non-unitary product transformations, within any given equivalence class of extremal PPT states with this rank. The first indication is simply based on parameter counting. As already discussed the number of equations that determine a product transformation that transforms a state of rank 4 to the projection form is 16 while the number of parameters to be determined by the equations is 17.

The second indication is found by listing the parameter sets $\{p_k, k=1,2,...,6\}$ for $SL$-equivalent states on projection form that we generate by our method. As shown in Fig.2a a large set of different distribution is found, consistent with the assumption that there is a continuum of equivalent states. If one of the parameters is restricted to a very limited interval, the corresponding sets $\{p_k\}$ seem either to be identical (up to the limitation set by the fixed parameter), or to divide into a small number of distinct groups, each of which are essentially identical. This is illustrated  in Fig.2b. This is consistent with the assumption that the full set is specified, up to a discrete set of possibilities, by a single continuous parameter.

The third indication is that we have checked numerically, in a few
examples, that if $\rho_0$ in \pref{transform2} is already on projection form,
then this equation has solutions of the form
$S=(\iden_A+\epsilon L_A)\otimes(\iden_B+\epsilon L_B)$ with $\epsilon$ an
infinitesimal parameter.  We find in each case that this set of
infinitesimal product transformations is uniquely determined if
we require $S$ to be hermitian and not unitary. 

All this seems to show that the picture is similar to that of the icosahedron case, so that any extremal PPT state of rank $(4,4)$ is $SL$-equivalent to a one parameter set of states on the projection form given by \pref{genprojec} and \pref{qpro}. And for each of these the partial transposed density matrices has the same projection form. (Note that the trivial equivalence under unitary product transformations is not included in this parameter counting.)

\begin{figure}[h]
\begin{center}
\includegraphics[width=16cm]{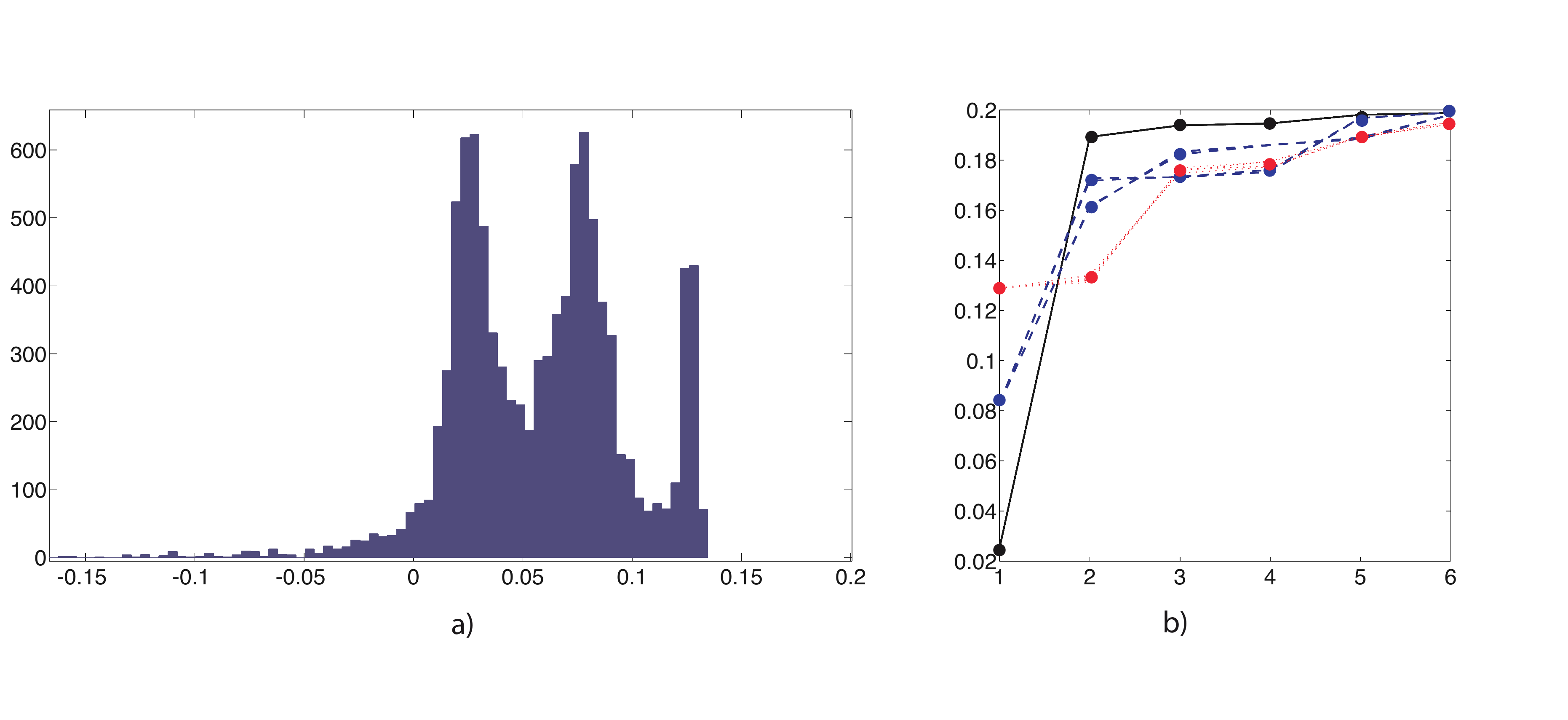}
\end{center}
\caption{\small {\bf a)} The figure shows the distribution of the smallest coefficient in the set $\{p_k\}$ for approximately 9500 $SL$-equivalent states on projection form, all of which are $SL$-equivalent to a generic rank $(4,4)$ extremal state. We observe that we find a large distribution that seems to cover all possible values from 0 up to an upper limit which is smaller than $1/6$, which is the upper limit of the icosahedron states. This implies that such generic rank $(4,4)$ extremal states are less symmetric than the icosahedron-class of states. The continuous nature of the distribution also supports the one-parameter family theory, while we do not have an explanation for the particular profile of the distribution. {\bf b)} This plot shows some of the sets $\{p_k\}$ where the smallest parameter is specified within three very limited intervals. With this limitation the rest of the set is found to be uniquely defined within essentially the same limitation. In one case, however (the dashed blue curve), there are two distinct groups of parameter sets with the same smallest coefficient.  These obervations support the claim that the $SL$-equivalent states on projection form defines a continuous one-parameter set of states.  \label{pk}}
\end{figure}

\section{Higher dimensions}
The lowest rank extremal PPT states in the $3\times 3$ system have their counterparts in higher dimensional systems.  Thus, in previous numerical studies of PPT states in several systems of dimension  $N_A\times N_B$, with $N_A$ and $N_B$ larger than $2$,  we have found such states with rank given by 
\be{eqrank}
r = N_A + N_B - 2
\ee
for both $\gr$ and $\gr^P$ \cite{LeinaasMyrheim10a}. These PPT states are typically extremal, all with a UPB with a finite number of (non-orthogonal) product vectors in their kernel. This is a situation which is different from what we find for entangled PPT states with other values of the rank. Based on these results  we have conjectured that, quite generally for higher dimensional systems, the number \pref{eqrank} gives the lowest rank of extremal PPT states {\em with full local ranks} for a bipartite system of dimension $N_A\times N_B$.  The number of product states for a UPB associated with any of these states is \cite{LeinaasMyrheim10a}
\be{pvecnum}
p = \frac{\left(N_A+N_B-2\right)!}{\left(N_A-1\right)!\left(N_B-1\right)!}
\ee
and the number of linearly independent product states is
\be{lustates}
d=N_A N_B-N_A-N_B+2
\ee
In Table 1 we have listed the relevant number of product states for the systems which we have studied numerically in this work.
The table also includes a list of dimensions for the sets of PPT states with the relevant ranks in these systems, as well as the number of parameters needed to specify the classes of $SL$-equivalent matrices of this type. These numbers are based on numerical studies that we have now performed, and which are described below. 

\begin{table}[h!]
\centering
\scalebox{0.85}{%
\begin{tabular}{|cccc|}
\hline
\multicolumn{1}{|c|}{\quad{\bf system}\quad  } & \multicolumn{1}{c|}{\quad\bf ranks} & \multicolumn{1}{c|}{\quad\bf p/d} & \multicolumn{1}{c|}{\quad\bf dimensions} \\
\hline\hline
$3\times 3$ & $(4,4)$ & $6/5$      & $36/4$ \\ \hline
$3\times 4$ & $(5,5)$ & $10/7$   & $55/9$ \\ \hline
$3\times 5$ & $(6,6)$ & $15/9$   & $77/13$ \\ \hline
$3\times 6$ & $(7,7)$ & $21/11$ & $102/16$  \\ \hline
$4\times 4$ & $(6,6)$ & $20/10$ & $75/15$ \\ \hline
$4\times 5$ & $(7,7)$ & $35/13$ & $96/18$  \\ \hline
$5\times 5$ & $(8,8)$ & $70/17$ & $119/23$ \\ \hline
\end{tabular}}
\caption{Ranks and numbers of product states in the kernel of lowest rank extremal PPT states (with full local rank) in a series of bipartite systems of low dimensions. The first column gives the dimensions of the two subystems, the second column the ranks of the density matrices and their partial transpose (which are equal for all these states), the third column gives the total number of product states in the kernel of the density matrices as well as the number of linearly independent product vectors, and the fourth column gives the dimension of the set of the density matrices and the number of parameters needed to parameterize the classes of $SL$-equivalent density matrices.} \label{tab}
\end{table}

The method used to determine the dimensions of the sets of PPT states is based on the counting of different ways to make small perturbations away from a density matrix $\gr$ of a given rank $(m,n)$, in such a way that the ranks of the matrix and its partial transpose are preserved. 
Consider then $\gr_0$ to be an extremal PPT state, and $(m,m)$ to be the (symmetric) ranks of the density matrix  and its transpose.  A perturbation of this state we write as 
\be{pert1}
\gr = \gr_0 + \epsilon\sigma, ~~~ \epsilon << 1
\ee
Further, let $P$ be the projector onto the image of $\gr_0$ and $Q$ the projector onto the image of $\gr_0^P$. The following conditions, $(I-P)\sigma(I-P) = 0$ and $(I-Q)\sigma^P(I-Q) = 0$, secures that  the ranks $(m,m)$, to first order in $\ge$, do not change under the perturbation. We rewrite the conditions as
\be{condpert1}
\sigma = P\sigma + \sigma P - P\sigma P\equiv \phi_P(\sigma), ~~~~~~ \sigma^P = Q\sigma^P + \sigma^PQ - Q\sigma^PQ\equiv \phi_Q(\sigma^P)
\ee
The maps $\phi_P$ and $\phi_Q$ define linear operators $\bs P$ and $\bs Q$ that act as projectors on the real vector space of hermitian matrices. The previous equations can therefore be written as
\be{condpert2}
\bs P\bs\sigma=\bs\sigma,~~~~~~ \bs Q\bs\sigma^P = \bs\sigma^P
\ee
with $\bs\sigma$ and $\bs\sigma^P$  viewed as vectors. The partial transpose can further  be expressed as a linear operator $\bs \Pi$ in this space, so that $\bs\Pi\,\bs\sigma=\bs\sigma^P$ \cite{LeinaasMyrheim07}. With the definition $\bar{\bs Q}\equiv \bs\Pi\bs Q\bs\Pi$,  the two equations in (\ref{condpert2}) can be combined in the single equation
\be{condpert3}
\bs P\bar{\bs Q}\bs P\bs\sigma = \bs\sigma
\ee
and the number of linearly independent matrices $\sigma$ that obey this equation will then be precisely the dimension of the set of matrices with rank $(m,m)$ in which $\gr_0$ sits.

The number of linearly independent matrices $\gs$ can be found by calculating the matrix $\bs P\bar{\bs Q}\bs P$ and counting the number of its eigenvalues that are equal to 1. As we are interested in the dimension of the set of {\em extremal} density matrices, with the specified rank, we have also checked whether the states close to $\gr_0$ are typically extremal. This has been done by repeatedly perturbing $\gr_0$ in different directions and checking the perturbed states for extremality. Since the rank, for a finite perturbation, will generally increase as a higher order effect in $\ge$, we have in each perturbation corrected for this by including a second search, if necessary, for a closeby density matrix with correct rank, before checking for extremality. 

The result is that for all the states listed in Table 1, we find that the density matrices with the correct ranks $(m,m)$ in the neighborhood of a chosen extremal density matrix are typically also extremal.  This we take as a clear indication that the dimension we calculate is also the dimension of the set of {\em extremal} states with the given rank.

To determine from this dimension the number of parameters that is needed to parametrize the classes of $SL$-equivalent states, we have subtracted the number of parameters that specify a product transformation of the form $SL(N_A,\complex)\times SL(N_B,\complex)$.  Each factor is specified by $2N^2-2$ real parameters, and the total number of parameters is therefore $2(N_A^2 + N_B^2) - 4$. The numbers given in Table 1 are obtained by such subtractions. In particular the number of parameters found in this way for the $(4,4)$ states of the $3\times 3$ system is $4$. This agrees with the conclusion reached in \cite{LeinaasMyrheim10b}.

\section{Projection operators in higher dimensions}

The same numerical methods that have been used to study the $(4,4)$ states  of the $3\times 3$ system we have also applied to the extremal PPT states in higher dimensional systems. 
A specific case is the $4\times 4$ system, where the relevant states are of rank $(6,6)$. These density matrices have a UPB with 20 product vectors in their kernel, 10 of these being linearly independent. As shown in Table 1 we find that these states form a 75-parameter subset within the 255-parameter set of density matrices of the $4\times4$ system. The number of parameters determining a product transformation in the $4\times4$ system is 60, which leaves us with 15 parameters with which to parameterize the equivalence classes. 

The method we use to search for PPT states of the correct rank and on projection form, works well also for this system, and we have by use of the method generated a large number of density matrices with these characteristics. We find, precisely as in the $3\times3$ system, that the density matrices always have a diagonal form, similar to \pref{qpro}, when expressed in terms of the product vectors of the UPB. In the $4\times4$ system the correct expressions for the density matrix $\gr$ and the corresponding projection $Q$  is
\be{projec4}
\gr={1\over 6}(\iden -Q)\,,\quad Q=10\sum_{k=1}^{20}p_k\psi_k\psi_k^\dag\,,\quad Q^2=Q
\ee
with $\psi_k=\phi_k\otimes \chi_k$ as the product states of the UPB. Precisely as in the $3\times3$ system we also here  find that the partially transposed operator $Q^P$ is a projection with the same rank as $Q$, so for the density matrices that we find there is complete symmetry between $\gr$ and $\gr^P$,
\be{projec4P}
\gr^P={1\over 6}(\iden -Q^P)\,,\quad Q^P=10\sum_{k=1}^{20}p_k\tilde\psi_k\tilde\psi_k^\dag\,,\quad (Q^P)^2=Q^P
\ee
where $\tilde\psi_k=\phi_k\otimes \chi_k^*$ are the product vectors of the conjugate UPB.

The states we find have generally $p_k\neq 0$ for all $k=1,2,...,20$. However, while most of these coefficients are positive, a small number of them will usually be negative. This is different from what we find in the $3\times 3$ system, where the typical situation is that all $p_k$ are positive, but where we occasionally find one of them to be negative.

Also for the $4\times4$ system we have investigated the possibility of transforming a generic extremal PPT state of rank $(6,6)$ to the projection form. The method we use is the same as for the $3\times3$ system, where we search for solutions to Eq.~\pref{transform2}, with $\gr_0$ as a density matrix with the correct rank, which is  generated by the method described in \cite{LeinaasMyrheim10a}. However, as opposed to the case with the $(4,4)$ states of the $3\times 3$ system, we have only for special choices of matrices $\gr_0$ been able to find product transformations that transform $\gr_0$ to the projection form. In the general case the iterative method that we use does not converge to an acceptable  solution. Instead it shows a slow convergence towards a singular transformation matrix.

The lack of convergence in this case may be a consequence of the increase in the number of variables in the problem, and therefore to a a decrease in the efficiency of the iterative procedure. But a clear possibility is that in the $4\times4$ system a generic extremal PPT state of rank $(6,6)$ cannot be transformed by a product transformation to the form of a projection. In fact, the form of the equation \pref{transform2} that we seek solutions for may indicate that this is the case. By counting the variables of the equation we find that the set of equations is underdetermined in the $3\times 3$ system, but it is overdetermined in the $4\times 4$ system as well as in other higher dimensional systems. (That does not, however, exclude the possibility that for the special density matrices that we consider there should exist solutions to the equation.)  

The other higher-dimensional systems that are listed in Table 1 have been studied by the same methods as the $3\times3$ and $4\times 4$ systems, and the results are essentially the same as for the $4\times 4$ system. This means that the searches for PPT states $\gr$ with rank $m$ as specified in Table 1, which are proportional to projection operators, are in most cases successful. By varying the initial value in the search we have therefore been able, for most of the listed systems, to generate a large number of different solutions. For systems of dimension $N=N_AN_B\geq20$ the iterative methods become rather slow, so the number of solutions we have found for them is somewhat smaller, though we find solutions also there.  In all cases we find density matrices of the same form as shown in \pref{projec4}. We also find in all cases that the partially transposed matrix $\gr^P$ is of the same rank as $\gr$. It is also a projection, and therefore we have a complete symmetry between $\gr$ and $\gr^P$, similar the one given by \pref{projec4} and \pref{projec4P} for the $4\times 4$ system.

For comparison we have also made searches for PPT states that are proportional to projections with ranks higher than those indicated in Table 1, in which case there is typically no UPB in the kernel of the matrix. The result is that we are able to find also such density matrices, but now the situation is different. In these cases the typical solution $\gr$ has a partial transpose $\gr^P$ with a higher rank and which is not proportional to a projection. 

We have for all the listed higher dimensional systems also performed searches for product transformations that transform a generic extremal PPT state, with the specified rank, to projection form. For these systems the result is the same as for the $4\times4$ system, that the searches in most cases are unsuccessful. Thus, only for the $3\times 3$ system do we find that we are able to transform the generic extremal PPT states of the given rank into the projection form. 

We summarize the main findings for the higher dimensional systems in the following way. For all the bipartite systems we have studied we have been able to generate entangled PPT states on projection form, with both $\gr$ and $\gr^P$ satisfying the conditions given by \pref{genprojec} and \pref{qpro}.  Whereas in the $3\times 3$ system all the entangled PPT states with the relevant rank seem to be $SL$ equivalent to the states on projection form, in higher dimensions these states seem instead to form a proper subset of the entangled PPT states with the given rank. In addition to these results we have made other observations that may be relevant for the generalized UPB construction, and we discuss some of them in the next section.

\section{States on projection form and real invariants}

We first return to a point briefly mentioned earlier in the paper about a difference in the UPB construction which appears when orthogonal UPBs are replaced by generalized UPBs. In the former case the PPT property follows directly from the projection form of $\gr$, since this implies also $\gr^P$ to be proportional to a projection, and therefore to be positive. With the use of a generalized UPB, we have no proof for this to be the case. Clearly the UPB construction will be simpler if we only have to take into account conditions that apply to $\gr$ and not to both $\gr$ and $\gr^P$, and for that reason we shall examine this point a bit further.

We first demonstrate that symmetry between $\gr$ and $\gr^P$ in the above respect is directly linked to a condition of equal ranks for the two matrices. At this point we make no specific assumptions about the matrix $\gr$ except for the normalization $\Tr[\gr]=1$. We refer to the rank of the matrix as $r$ and $P$ as a projection on the image of $\gr$. We have
\be{trace}
\Tr[(\gr-{P\over  r})^2]=Tr[\gr^2]-{1\over r}
\ee
which implies 
\be{trace2a}
Tr[\gr^2]\geq{1\over r}
\ee
with equality when $\gr=P/r$. We next assume precisely this to be the case, so that $\gr$ is proportional to a projection and therefore $Tr[\gr^2]=1/r$. We further introduce $s$ as the rank of $\gr^P$ and apply the same argument to this matrix, so that
\be{trace2b}
Tr[(\gr^P)^2]\geq{1\over s}
\ee
On the other hand, the trace of the matrix squared is invariant under the operation of partial transposition, $Tr[(\gr^P)^2]=Tr[\gr^2]=1/r$, and this implies the inequality
\be{ineq}
s\geq r
\ee
This means that the rank of $\gr^P$ is larger or equal to the rank of $\gr$ when the latter is proportional to a projection, and there is equality between the ranks if and only if $\gr^P$ is also proportional to a projection. From this we conclude that $\gr^P$ satisfies the same projection condition \pref{genprojec} 
as $\gr$ if and only if they have equal ranks.

As a particular application let us assume that $\gr$ is of the projection form given by \pref{genprojec} and \pref{qpro}, and that it can be transformed to a situation where the product vectors $\psi_k=\gf_k\otimes\chi_k$ of the generalized UPB are decomposed into vectors $\gf_k$ and $\chi_k$ of the subsystems where the vectors of one of these sets are real. This may be taken as subsystem B, so that $\chi_k^*=\chi_k$. As a consequence the density matrix is invariant under partial transposition, and therefore the ranks of $\gr$ and $\gr^P$ are equal, not only after the transformation, but also in the reference frame where $\gr$ is on projection form. This is so since the rank is invariant under the non-singular product transformations. As follows from the discussion above the partial transpose $\gr^P$ will in the same reference frame also be proportional to a projection, and hence positive.

The condition of real vectors is an obvious way to secure the density matrix and its partial transpose to have the same rank. On the other hand, this condition could possibly be too restrictive to make it interesting. However, in 
\cite{LeinaasMyrheim10b} a set of $SL\otimes SL$ invariants was introduced that characterize the product vectors of the generalized UPBs, and for the relevant rank $(4,4)$ states of the $3\times 3$ system they were found to be real. This is a non-trivial result, since for a UPB more generally the invariants will be complex.  When all the invariants are real it follows that the vectors of the UPB can be transformed to real form. 
For the higher dimensional systems the situation is more complex. The $3\times n$ systems ($n> 3$) that are available for numerical study we have found all to have real invariants for the set of vectors associated with one of the subsystems, but complex for the other subsystem. However, for other systems, like the $4\times 4$ system, the generic lowest rank extremal PPT states are characterized by $SL$ invariants that are complex for both subsystems.

A similar study of the states referred to in the present work, where the density matrices are on projection form, however shows a somewhat different picture. For all the systems we have examined we find the invariants of the states that have been generated are all real for at least one of the subsets of vectors $\{\gf_k\}$ or $\{\chi_k\}$. Thus, in all these cases one of the sets of vectors can be transformed to real form. This indicates a close connection between the reality condition referred to above and the condition of projection form for $\gr$ and $\gr^P$. Note however, that to impose the reality condition and the projection condition simultaneously may be too restrictive, since the conclusion we can draw from our studies is only that the density matrices we have found, which satisfy the projection condition, can be {\em transformed} to a form where one of the subsets $\{\phi_k\}$ or $\{\chi_k\}$ consists only of real vectors.

\section{Conclusions}

The motivation for the present work has been to examine the possibility of generalizing the method of constructing entangled PPT states with the use of unextendible product bases (UPB). The established form of this construction is to use a set of orthogonal product states, with no product states in the orthogonal subspace, and to define the corresponding density matrix as a projection operator with the UPB in its kernel. The method applies particularly to rank $(4,4)$ states in bipartite quantum systems of dimension $3\times3$, and as previously shown in numerical studies all entangled PPT states with this rank seem to be equivalent under product transformations to the states constructed in this way \cite{LeinaasMyrheim10b}.

In higher dimensional bipartite systems there are states that share many of the properties with the $(4,4)$ states of the $3\times3$ system. They have all a finite and complete set of product vectors in their kernel and no product vector in their image, and they all seem also to share  the property of being the lowest rank extremal PPT states in the system under consideration. The idea is to extend the UPB construction to these states. It seems not to be possible to extend the construction with {\em orthogonal} product vectors to higher dimensions, and we have focussed on the possibility of using non-orthogonal UPBs which keep some of the other properties of the original construction. Our assumption is that we can define these density matrices more generally as projections which can be expressed in a particular way in terms of product vectors of the non-orthogonal UPB. 

We have first examined this generalization for rank $(4,4)$ extremal PPT states of the $3\times 3$ system. In this case an explicit example can be found, with the vectors of the generalized UPB defined by the symmetry axes of an icosahedron. All the six product vectors of this UPB appear with equal weight in the definition of the corresponding extremal PPT state. 
We can further show, by a linear deformation of the icosahedron, that a one-parameter set of different density matrices on projection form exists, where the matrices are equivalent under product transformations. 

To study more general states we have applied numerical methods. The first method is to search for PPT states which have rank 4 and which are proportional to projections. The searches have been used to generate a large set of such states and these states have been found always to be on the form suggested by the generalized UPB construction, where $\gr$ and $\gr^P$ have equal ranks, are both proportional to projections and have the same form when expressed in terms of the product vectors of the UPBs associated with the two matrices.

A second method has been used to check whether generic extremal PPT states of rank $(4,4)$ are equivalent under product transformations to density matrices of the form suggested for the generalized UPB construction. This has been demonstrated numerically and the numerical studies suggest that each state is equivalent to a one-parameter set of density matrices of this form where the matrices associated with an orthogonal UPB constitute a discrete subset.

To examine the relevance of the generalized UPB construction in bipartite systems of higher dimensions, we have first studied numerically the dimensions of the sets of extremal PPT states of the corresponding ranks. We  have then applied the same numerical methods as used for the $3\times3$ system to search for density matrices of the suggested form. The search is thus for PPT states on projection form, with the specified rank, and the result are similar to those found in the $3\times 3$ system. For all the systems a large number of states which satisfy the search criteria have been found, and they all show a complete symmetry between the density matrix and its partial transpose. They have the same rank, are both projections and can be expressed in terms of the product vectors of the associated UPBs in the same way. This demonstrates that the requirements suggested for a generalized UPB construction  are satisfied for a large set of states in all these systems.

We have however observed a difference between the $3\times 3$ system and higher-dimensional  systems when performing searches for states of projection form that are equivalent under product transformations to a randomly generated extremal PPT state of the correct rank. As we mentioned, in the $3\times3$ system the results indicate that any  extremal PPT state with the specified rank is equivalent to a one-parameter set of states on projection form, where the states associated with {\em orthogonal} UPBs form a discrete subset. In higher dimensions similar searches have been unsuccessful in most cases. This may suggest that the density matrices in higher dimensions that are of the form specified in the searches form a proper subset of the full set of extremal PPT states with the given rank. This is further supported by the fact that the states on projection form of the correct rank in higher dimensions are found to have real $SL$-invariants, in contrast to the all complex invariants of randomly generated states of the same rank.

Based on these results we suggest that the generalized UPB construction may be relevant for construction of low-rank extremal PPT states in higher dimensions.
Our study thus shows that extremal PPT states of the suggested form exist in all the higher-dimensional systems we have been able to examine. However, we still lack a 
concrete prescription for constructing generalized UPBs which define density matrices with the right properties. This problem, to find such a prescription, and also the question about how general the density matrices of this form are, we shall therefore have to leave as interesting questions for further studies.

\section{Acknowledgements}
Financial support from the Norwegian Research Council is gratefully acknowledged.


\end{document}